\documentclass[aps,reprint,superscriptaddress,showpacs]{revtex4-1}
\usepackage{graphicx}
\usepackage{epsfig}
\usepackage{epstopdf}
\usepackage{dcolumn}
\usepackage{bm}

\begin{document}


\title{Single-shot readout of a superconducting flux qubit with a flux-driven Josephson parametric amplifier}

\author{Z.~R. Lin}
\affiliation{RIKEN Center for Emergent Matter Science, Wako, Saitama 351-0198, Japan}

\author{K. Inomata} 
\affiliation{RIKEN Center for Emergent Matter Science, Wako, Saitama 351-0198, Japan}

\author{W. D. Oliver}
\affiliation{MIT Lincoln Laboratory, 244 Wood Street, Lexington, Massachusetts 02420, USA}

\author{K. Koshino}
\affiliation{College of Liberal Arts and Sciences, Tokyo Medical and Dental University, 2-8-30 Konodai, Ichikawa 272-0827, Japan} 

\author{Y. Nakamura}
\affiliation{RIKEN Center for Emergent Matter Science, Wako, Saitama 351-0198, Japan}
\affiliation{Research Center for Advanced Science and Technology, The University of Tokyo, Meguro-ku, Tokyo 153-8904, Japan}

\author{J. S. Tsai}
\affiliation{RIKEN Center for Emergent Matter Science, Wako, Saitama 351-0198, Japan}
\affiliation{NEC Smart Energy Research Laboratories, Tsukuba, Ibaraki 305-8501, Japan}

\author{T. Yamamoto} \email[t-yamamoto@fe.jp.nec.com]{}
\affiliation{RIKEN Center for Emergent Matter Science, Wako, Saitama 351-0198, Japan}
\affiliation{NEC Smart Energy Research Laboratories, Tsukuba, Ibaraki 305-8501, Japan}

\date{\today}

\begin{abstract}
We report single-shot readout of a superconducting flux qubit 
by using a flux-driven Josephson parametric amplifier (JPA). 
After optimizing the readout power, gain of the JPA and timing of the data acquisition, 
we observe the Rabi oscillations with a contrast of 74\% which is mainly limited by 
the bandwidth of the JPA and the energy relaxation of the qubit. 
The observation of quantum jumps between the qubit eigenstates 
under continuous monitoring indicates 
the nondestructiveness of the readout scheme.
\end{abstract}


\maketitle

A dispersive readout scheme using an electromagnetic resonator is widely used 
to read out the state of superconducting qubits.~\cite{Devoret13}  
This technique utilizes the fact that the resonance frequency of a resonator, 
to which the qubit is dispersively coupled, 
depends on the state of the qubit.~\cite{Blais04} 
Its intriguing properties, e.g., high fidelity and nondestructiveness, 
have been demonstrated in a variety of experiments exploring cavity 
quantum electrodynamics using superconducting circuits.~\cite{Wallraff05, Schuster07} 
One of the problems with dispersive readout, however, 
is its low signal to noise ratio (SNR). 
In order to minimize backaction on the qubit, 
one has to detect a small microwave signal leaking out of the 
resonator containing on the order of a single photon, 
and this should be done within a time much shorter than the lifetime of the qubit. 
Therefore, low-noise and wide-band amplifiers are indispensable. 
To date, cryogenic HEMT amplifiers have been most commonly used. 
However, even with a state-of-the-art HEMT amplifier
with a noise temperature of $\sim$ 2-5~K, 
the SNR typically remains smaller than unity. 
This necessitates averaging the readout signal 
over many trials (identical realizations of the experiment) to discriminate the 
state of the qubit, whereas single-shot readout is desirable 
for applications in quantum information processing. 

Recently, many groups have attacked this problem by using 
low-noise amplifiers based on superconducting 
circuits,~\cite{Siddiqi06,Lupascu07,Mallet09,Johnson11} 
including so-called Josephson parametric amplifiers 
(JPAs),~\cite{Beltran07,Yamamoto08,Bergeal10} 
and single-shot readout using JPAs has been realized.~\cite{Vijay11,Johnson12,Riste12}  
The JPA we have developed is pumped by ac flux instead of ac current, 
and we call it the ``flux-driven JPA".~\cite{Yamamoto08} 
It can be operated in the degenerate three-photon mode 
(signal and pump frequencies are related by a factor two)
and has the practical advantage that it does not require an additional microwave field to 
cancel the pump leaking out of the device. 
Recently, the vacuum noise squeezing has been demonstrated using the same type of the device.~\cite{Menzel12,Zhong13}
Also, similar device based on the lumped-element circuit has been reported.~\cite{Mutus13}
We note that there is another type of JPA called the Josephson ring modulator, 
which may be also operated in a three-photon mode, but in a nondegenerate configuration.~\cite{Bergeal10,Abdo11} 

In this letter, we utilize a flux-driven JPA as a preamplifier 
for the dispersive readout of a superconducting flux qubit.~\cite{Inomata12}
We observed a drastic improvement in the SNR, and this enabled us to achieve 
single-shot readout, which means that we judge the state of the qubit
by applying a single readout pulse.
We observed Rabi oscillations with a contrast of 74\% 
which is limited mainly by the bandwidth of the JPA and the energy relaxation of the qubit. 
Also, by continuously monitoring the qubit, 
we observed quantum jumps between the qubit eigenstates.~\cite{Vijay11} 
The statistics of the jumps confirm the nondestructiveness of our readout.

\begin{figure}
\includegraphics[width=0.9\columnwidth,clip]{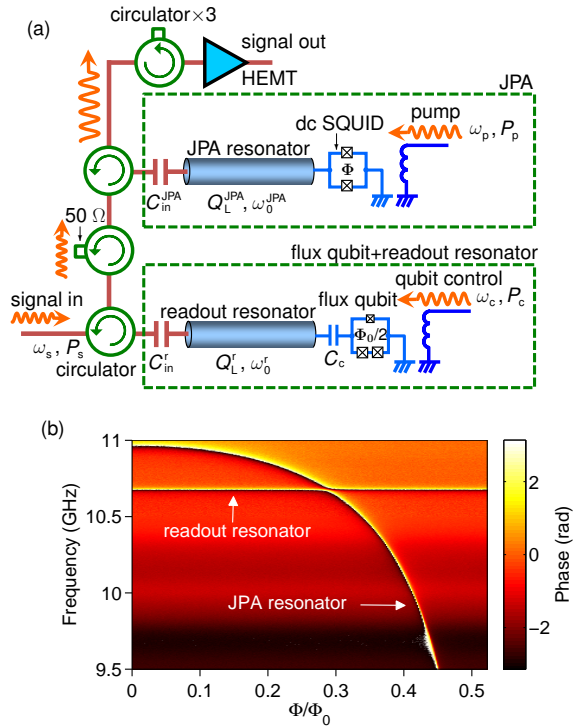}
\caption{~(Color online) (a) Schematic of the measurement setup. 
The designed value of $C_{\rm in}^{\rm JPA}$, $C_{\rm in}^{\rm r}$ and $C_{\rm c}$ are 
16~fF, 16~fF and 4~fF, respectively. 
$Q_{\rm L}^{\rm JPA}$ and $Q_{\rm L}^{\rm r}$ are measured to be 310 and 650, respectively. 
(b) Phase response of the circuit as a function of the input signal frequency 
and the flux bias for JPA.~\label{fig1}
}
\end{figure}

Figure~\ref{fig1}(a) shows our measurement setup for the single-shot readout of 
a flux qubit using flux-driven JPA. 
All measurements were performed using a dilution refrigerator at the 
base temperature T $\sim 10$~mK.
The setup is the same as that used in Ref.~\onlinecite{Inomata12}, 
with the addition of a flux-driven JPA contained in a separate sample holder 
together with an independent coil for the flux bias $\Phi$. 
The chip containing the qubit is the same one used in Ref.~\onlinecite{Inomata12}, 
in which a conventional three-Josephson-junction flux qubit is capacitively 
coupled to a coplanar waveguide (CPW) resonator with 
a loaded quality factor $Q^{\rm r}_{\rm L}$ of 650. 
We refer to this resonator as the ``readout resonator" to distinguish it from the ``JPA resonator". 
In the present study, we bias the qubit at $\Phi=0.5\Phi_0$, 
where $|0\rangle$ to $|1\rangle$ transition frequency $\omega_{01} = 2\pi \times 5.461$~GHz 
is insensitive to low frequency flux noise to first order in the flux fluctuation $\delta \Phi$. 
The resonant frequency of the resonator $\omega_0^{\rm r}$ is $2\pi \times 10.674$~GHz 
when the qubit is in the ground state. 
The dispersive shift of the resonator, $2\chi$, namely the shift in $\omega_0^{\rm r}$ 
when the qubit is excited from ground state $|0 \rangle$ to excited state $|1 \rangle$,
is enhanced by the the straddling effect,~\cite{Inomata12} 
and measured to be $2 \chi = 2\pi \times (-80)$~MHz. 

The JPA used in this study has the same design as that reported 
in Ref.~\onlinecite{Yamamoto08} and was fabricated by the same process. 
The device consists of a CPW resonator with a SQUID termination and 
a pump line inductively coupled to the SQUID loop. 
The critical current of each Josephson junction in the SQUID was estimated to
be 1.2~$\mu$A from the process test data. 
The dc flux bias through the loop determines the static resonant frequency $\omega_0^{\rm JPA}$, 
namely the band center of the JPA. A microwave field applied to the pump port at frequency $2\omega_0^{\rm JPA}$ 
does parametric work on an incoming microwave field at the signal port with frequency around $\omega_0^{\rm JPA}$. 
Because the pump and the signal frequencies differ by a factor two, 
we do not need an additional microwave field to cancel the pump in the output line,
as  is often used for current-driven JPAs.~\cite{Riste12,Mallet11} 

Figure~\ref{fig1}(b) plots the phase response of the 
signal transmission through the circuit shown in Fig.~\ref{fig1}(a) 
as a function of the input signal frequency $\omega_{\rm s}$ and 
the JPA flux bias $\Phi$. The response was measured by a vector network analyzer, 
with the JPA pump turned off. At each $\Phi$, we observe two resonances. 
The one at 10.674~GHz which is independent of $\Phi$ is due to the readout resonator, 
while the other which depends on $\Phi$ is due to the JPA resonator. 
They become equal at $\Phi/\Phi_0=0.29$, and this is the bias point 
where we operate the JPA.~\cite{comment1} 
The external and internal quality factors of the JPA resonator measured 
at slightly detuned $\omega_0^{\rm JPA}$ of 10.803~GHz are $330$ and $4900$, respectively. 

\begin{figure}
\includegraphics[width=0.9\columnwidth,clip]{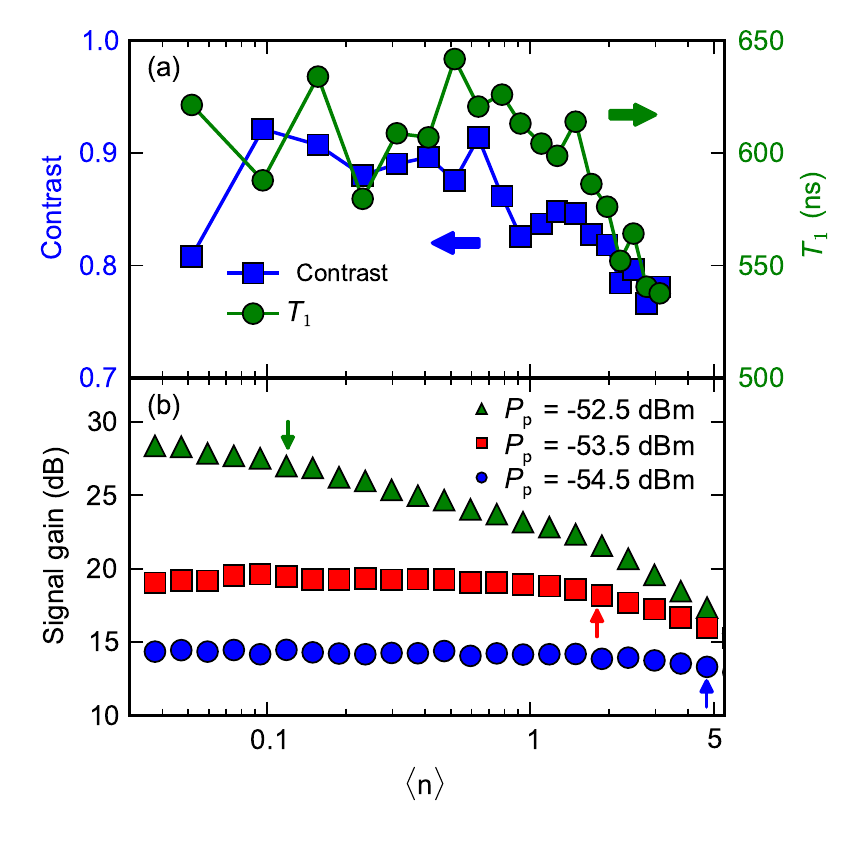}
\caption{~(Color online) (a) $T_1$ and contrast of Rabi oscillations as functions of 
the number of readout photons $\langle n \rangle$. 
For the $T_1$ measurement, 
a readout pulse of a length $3~\mu$s 
is applied after the $\pi$-pulse.
We acquire readout data of duration 100 ns, at a repetition rate 167
kHz, and average over 65,000 such trials. 
(b) Dependence of the signal gain on $\langle n \rangle$ for different 
pump powers. The arrows indicate the 1-dB-compression points. 
The JPA is operated in the nondegenerate mode, namely, 
at $\omega_{\rm p} = 2\omega_{\rm s} - 2\pi \times 100$~kHz, where 
$\omega_{\rm s} = \omega_0^{\rm r}$. 
\label{fig2}
}
\end{figure}

\begin{figure}
\includegraphics[width=0.9\columnwidth,clip]{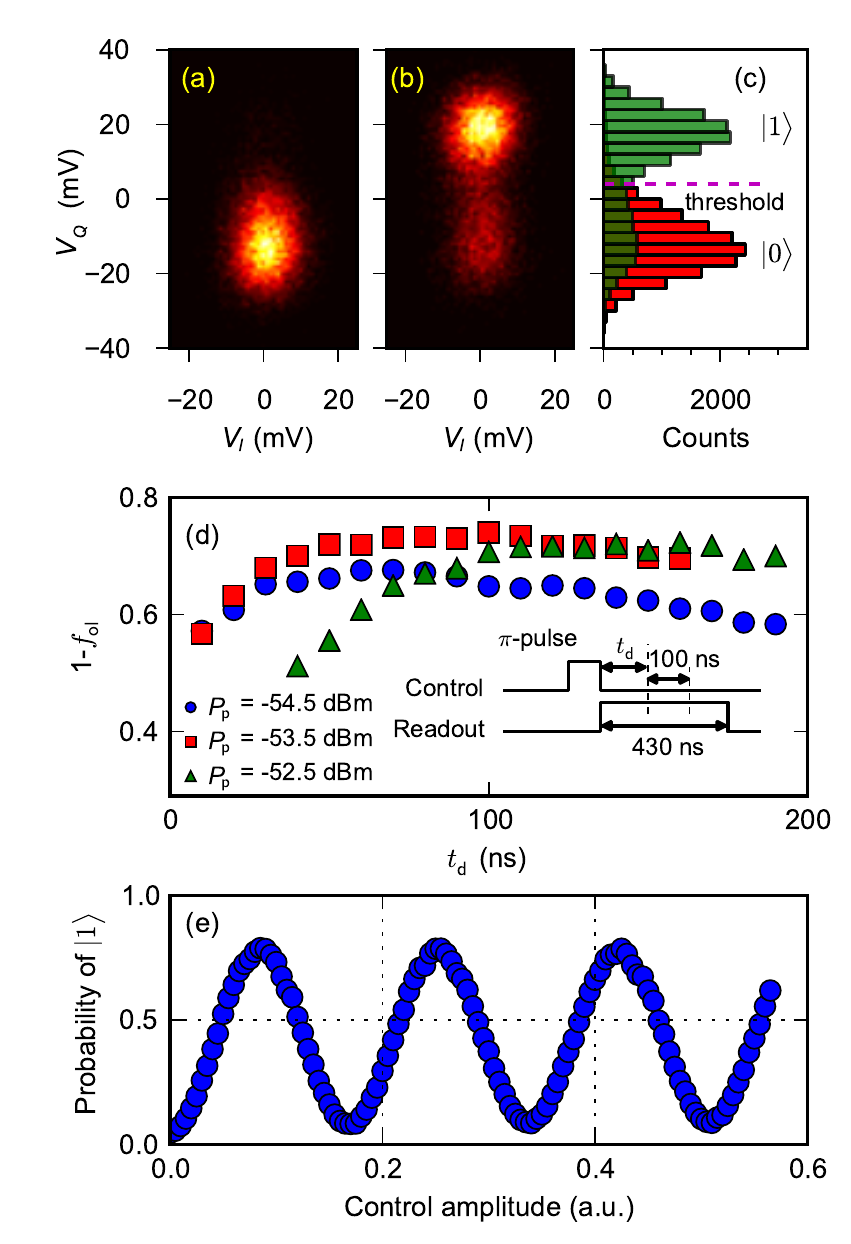}
\caption{~(Color online) 
Histograms of the digitally down-converted IF signal in the $I$--$Q$ plane
(a) without and (b)~with a $\pi$ pulse. 
Both figures are rotated around the origin by a phase offset 
so that the peaks of the distribution lie along the $V_Q$ axis. 
The JPA is operated at a gain of 25.6~dB and a bandwidth of 4.8~MHz 
($P_{\rm p} =-53.5$~dBm), and $t_{\rm d}$ is set to be 100~ns. 
(c) Histograms obtained from (a) and (b) projected onto the $V_Q$ axis. 
The horizontal dashed line represents the threshold used to discriminate the qubit states. 
(d) Contrast ($1-f_{\rm ol}$) as a function of $t_{\rm d}$.
All data were measured with $\langle n \rangle = 1.3$. 
(e) Rabi oscillations observed with averaged single-shot measurements.~\label{fig3} 
}
\end{figure}

First, we optimize the signal power for the readout. 
A large signal power is favorable in terms of SNR, but it can induce qubit energy relaxation 
due to readout backaction.~\cite{Boissonneault09,Slichter12} 
In Fig.~\ref{fig2}(a), we show the qubit energy relaxation time $T_1$ 
as a function of the number of readout photons $\langle n \rangle$ 
in the readout resonator. 
Here $\langle n \rangle$ is defined as $4P_{\rm s}Q^{\rm r}_{\rm L}/ \hbar \omega_{\rm s}^2$, 
where $P_{\rm s}$ is the power of the readout microwave field 
at the input of the readout resonator.~\cite{Clerk10} 
In this measurement, the pump of JPA is turned off, and 
$\omega_0^{\rm JPA}$ is far detuned from the signal frequency. 
The qubit relaxation time $T_1$ is determined from the decay of an ensemble 
averaged signal over a number of repetitions of the qubit excited by a 
$\pi$-pulse and a subsequent continuous readout. 
We observe a decrease of $T_1$ when $\langle n \rangle$ $\gtrsim$ 2. 
In the figure, we also plot the contrast of Rabi oscillations, 
which was obtained from ensemble average measurement using a pulse sequence similar to the one shown in the inset of 
Fig.~\ref{fig3}(d) with an optimized delay time $t_{\rm d}$.~\cite{Inomata12} 
Because the contrast is mainly limited by $T_1$ here,
it exhibits a similar dependence on $\langle n \rangle$ to that of $T_1$. 
These data show that $\langle n \rangle$ should not be larger 
than $\sim$ 2 to guarantee the nondestructiveness of the readout.

We also checked the dynamic range of the JPA. 
Figure~\ref{fig2}(b) shows the dependence of the signal gain 
on $\langle n \rangle$ when $P_{\rm p}$ is
$-52.5$~dBm, $-53.5$~dBm and $-54.5$~dBm. 
We observed unwanted parametric oscillations when $P_{\rm p} \ge -52.0$~dBm. 
The measurement was done using a continuous microwave and a spectrum analyzer. 
The signal gain is measured in a nondegenerate mode configuration, namely 
at a pump frequency $\omega_{\rm p} = 2\omega_{\rm s} - 2\pi \times 100$~kHz, 
where $\omega_{\rm s} = \omega_0^{\rm r}= 2\pi \times 10.674$~GHz. 
The 1-dB-compression points for these $P_{\rm p}$'s are indicated 
by the arrows in the figure.

Next, we tried to maximize the single-shot readout contrast. 
We operate the JPA in a degenerate mode configuration, i.e.,
$\omega_{\rm p} = 2\omega_{\rm s} = 2\omega_0^{\rm r}$. 
The phase of the pump microwave is adjusted so that the phase-dependent gain becomes maximal. 
Using the pulse sequence shown in the inset of Fig.~\ref{fig3}(d), 
we acquired the reflected readout pulse in the time domain using a heterodyne detection scheme
with with an IF frequency of 50 MHz and a sampling rate of 1 GS/s. 
We used a $100$-ns-long data sequence, the acquisition of which begins after a delay
$t_{\rm d}$ from the end of the applied $\pi$-pulse, to extract
the amplitude and phase of the reflected readout pulse. 
The 100-ns-long data acquisition is 
needed because of the limited SNR;
the noise consists mainly of the input vacuum noise and possibly the noise added by JPA. 
From the SNR, we can estimate 
the  upper limit of the noise temperature of the JPA, which is found to be $380$ mK.~\cite{suppl}
More precise measurement using a calibrated noise source has recently been reported 
using the same kind of device.~\cite{Zhong13} 
Figures~\ref{fig3}(a) and (b) show examples of 
histograms for $1.6\times 10^4$ such measurements 
without and with the $\pi$ pulse, respectively. 
As shown in Fig.~\ref{fig3}(c), 
which are the histograms obtained from Figs.~\ref{fig3}(a) and (b) projected onto the $V_Q$ axis, 
we observe two, well-separated distribution peaks which correspond to 
the qubit $|0 \rangle$ and $|1 \rangle$ states. 
From such data, we calculate the overlap between the two peaks, 
defined by 
\begin{equation}
f_{\rm ol} = \sum_i \min(H_0(i),H_1(i))/N_{\rm tot}, 
\end{equation}
where $H_0$ ($H_1$) is the histogram when the $\pi$ pulse is turned off (on), 
$i$ is the index for the bin of the histogram, 
and $N_{\rm tot}$ is the total number of counts for each histogram. 
In Fig.~\ref{fig3}(d), we plot single shot readout contrast defined 
by $1-f_{\rm ol}$ as a function of $t_{\rm d}$ 
 for the $P_{\rm p}$'s used in Fig.~\ref{fig2}(b). 
All data were measured with $\langle n \rangle = 1.3$. 
Qualitatively, the contrast at a fixed $P_{\rm p}$ is determined by 
the competition between energy relaxation of the qubit and 
the ring-up of the readout signal whose timescale is determined by the bandwidth of the JPA. 
Namely, a short $t_{\rm d}$ is favorable for high contrast 
in terms of small probability of energy relaxation, 
but unfavorable in terms of state distinguishability. 
As a consequence, there is an optimal $t_{\rm d}$ for each $P_{\rm p}$ 
as shown in Fig. 3(d), although the drop of the contrast due to energy relaxation 
for $P_{\rm p}=-52.5$~dBm is not clearly observed for $t_{\rm d} \le 190$~ns. 
Among these three $P_{\rm p}$'s, we attained the maximum contrast with 
$P_{\rm p}=-53.5$~dBm and $t_{\rm d}\sim100$~ns. 
The histogram shown in Fig.~\ref{fig3}(c) is obtained with this condition. 
Using the same condition, we observe Rabi oscillations as shown in Fig.~\ref{fig3}(e). 
Here, the probability of the qubit in state $|1 \rangle$ is plotted as a function 
of the amplitude of the control pulse of duration 50~ns. 
To discriminate the qubit state, we set a threshold as shown in Fig.~\ref{fig3}(c).
The attained contrast is 73.5\%. 
Possible sources of the error are (i)~incomplete initialization of the qubit, 
(ii)~insufficient separation of the peaks in the histogram, 
and (iii)~qubit energy relaxation. 
Errors from each source are estimated in the following.~\cite{suppl}
The error from the first source (incomplete initialization) is estimated from the histogram 
when no $\pi$-pulse is applied. This histogram was obtained with a higher JPA gain of 29.0~dB 
and longer $t_{\rm d}$ of 300~ns to clearly separate out the error counts. 
This error is estimated to be 2.8\%. 
The error from the second source (insufficient peak separation) is estimated from the overlap 
between the two Gaussian functions fitted to the histogram peaks 
corresponding to the qubit states $|0\rangle$ and $|1\rangle$, and calculated to be 2.4\%. 
The error from the third source (energy relaxation) is estimated from 
the histogram when a $\pi$ pulse is applied. 
In the histogram, we observe two main peaks which correspond to the qubit state $|1\rangle$ 
and the qubit state $|0\rangle$. 
The latter is due to the qubit energy relaxation, 
and the peak area (14.8\%) roughly agrees with the value expected from the 
exponential decay during $t_{\rm d}$. 
In addition, we observe counts in the middle of those peaks, 
more than one would expect from the Gaussian curves fitted to these peaks. 
We attribute them to the relaxation during our acquisition time of 100~ns. 
In total, we estimate the error due to the qubit relaxation to be 19.8\%. 
We still have an unaccounted error of 1.5\%, 
which could be related to the transient response of JPA. 

\begin{figure}
\includegraphics[width=0.9\columnwidth,clip]{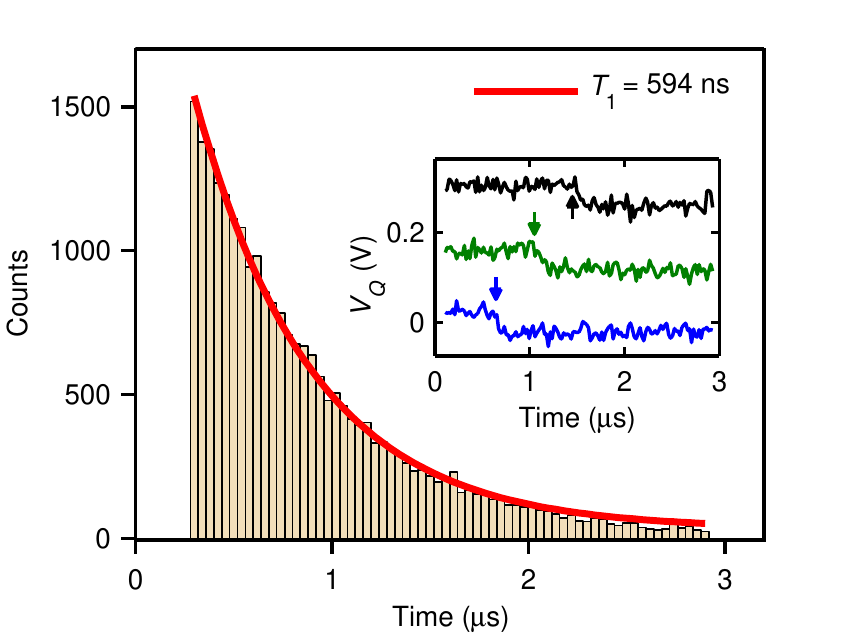}
\caption{~(Color online) 
Histogram showing the statistical distribution of jump times of the qubit 
from state $|1\rangle$ to $|0\rangle$. 
The solid curve is an exponential fit with a time constant of 594~ns. 
The JPA is operated at a gain of 29.0~dB and a bandwidth of 2.8~MHz. 
The readout photon number $\langle n \rangle$ is 1.3. 
Inset: Examples of the time trace under continuous monitoring. 
Arrows indicate the occurrence of quantum jumps.~\label{fig4}
}
\end{figure}

To check the nondestructiveness of our readout, 
we performed continuous monitoring of the qubit.~\cite{Vijay11}
After preparing the qubit in state $|1\rangle$ using a $\pi$-pulse, 
we applied a 3-$\mu$s-long readout pulse 
and extracted the amplitude and phase using a 40~ns-long time trace initiated every 20~ns. 
To clearly discriminate the qubit states with shorter integration time, 
we set $P_{\rm p}=-52.5$~dBm to improve the SNR, 
where the JPA exhibits a gain of 29.0~dB and a bandwidth of 2.8~MHz. 
The inset of Fig.~\ref{fig4} shows examples of the time trace of $V_Q$, 
where we can clearly observe jumps as indicated by the arrows. 
From such time traces, we extract the jump time of the qubit from 
state $|1\rangle$ to $|0\rangle$ by setting a threshold in $V_Q$. 
Fig.~\ref{fig4} shows the statistical distribution of the jump time. 
Due to the JPA's limited bandwidth of 2.8~MHz, 
which corresponds to a time constant of $\tau=114$~ns, 
we cannot correctly detect the jumps earlier than $\sim \tau$. 
Thus, we fit the histogram from 300~ns to $2.9~\mu$s with an exponential function
to obtain the decay time constant.
The time constant is 594~ns, which is consistent with $T_1$ of 600~ns 
measured using a standard pulse sequence, namely, 
a $\pi$-pulse followed by a delayed readout pulse. 
This consistency indicates that our readout is highly nondestructive. 

In conclusion, we demonstrated single-shot readout of 
a superconducting flux qubit with a flux-driven JPA. 
The flux-driven JPA has a widely tunable band center and is 
easily operated in the degenerate three-photon mode. 
With the aid of the flux-driven JPA, we achieve a single-shot readout 
contrast of 74\%, which is currently limited by the bandwidth of the JPA 
and the qubit energy relaxation time. 
We also observed quantum jumps of the qubit, which indicates that 
our readout is highly nondestructive. 
This measurement technique can be further applied to quantum feedback control,~\cite{Vijay12} 
and single microwave photon detection.~\cite{Koshinoinp}

\begin{acknowledgments}
The authors are grateful to V. Bolkhovsky and G. Fitch for
assistance with the device fabrication at MIT-LL.
This work was partly supported 
by the Funding Program for World-Leading Innovative R\&D 
on Science and Technology (FIRST), Project for Developing 
Innovation Systems of MEXT, MEXT KAKENHI (Grant 
Nos. 21102002, 25400417), SCOPE (111507004), and National 
Institute of Information and Communications Technology (NICT).
\end{acknowledgments}

\appendix
\section{The error budget of the Rabi contrast}
In the main article, we showed Rabi oscillations with a contrast of 73.5\%. 
Here, we present our analysis on the loss of the contrast. 
Possible errors are from incomplete initialization ($\alpha$), 
insufficient separation of the peaks in the histogram ($\beta$), 
and energy relaxation of the qubit ($\gamma$). 
In the following, we explain these errors one by one. 

\subsection{initialization error}
\begin{figure*}
\begin{center}
\includegraphics[width=1.3\columnwidth,clip]{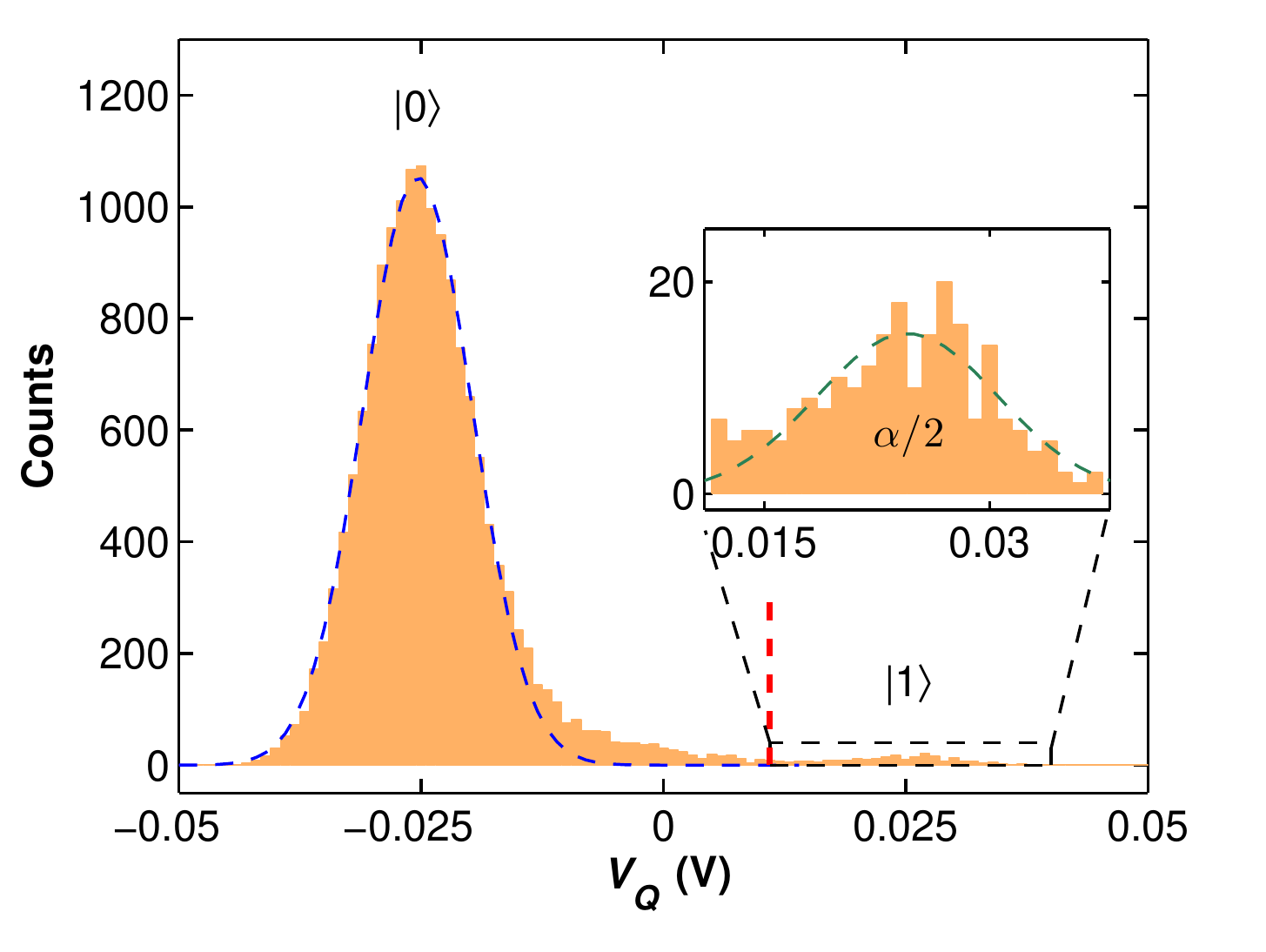}
\end{center}
\caption{Histogram when the $\pi$ pulse is turned off.
The JPA is operated with a gain of 29.0~dB and a bandwidth of 2.8~MHz, 
and the delay time $t_{\rm d}$ is 300~ns. 
The blue dashed curve is a Gaussian fit to the peak corresponding 
to the qubit state $|0\rangle$. 
The red dashed line is the threshold used to discriminate the qubit states. 
The inset shows the magnification of the part indicated by the black dashed box. 
The green dashed curve in the inset is a Gaussian fit to the $|1\rangle$-state peak. 
}
\label{figS1}
\end{figure*}

The qubit is nominally initialized to the ground state $|0\rangle$ by 
waiting a sufficiently
long time for thermalization to occur, with the assumption that the qubit
temperature (i.e., $k_B T$) is smaller than the  $|0\rangle$ -  $|1\rangle$ 
level separation (i.e., $\hbar\omega_{01}$).
Figure~\ref{figS1} shows the histogram when no $\pi$ pulse is applied. 
In order to increase the separation between the two distribution peaks 
corresponding to the qubit states $|0\rangle$ and $|1\rangle$, 
the gain of JPA is increased to 29.0~dB and the delay time $t_{\rm d}$ is increased to 300~ns, 
compared to the condition used for Rabi oscillation measurement shown in the 
main article. 

As seen in the figure, 
we observe non-zero population of the qubit in state $|1\rangle$.
The green dashed curve in the inset is a Gaussian fit to the $|1\rangle$-state peak.
From the fitting, we estimate the 
the $|1\rangle$ state population to be 1.4\%, 
which corresponds to the qubit temperature of $\sim$60~mK. 
The corresponding loss in Rabi contrast is twice this value.
Thus, we estimate $\alpha$ to be 2.8\%. 

\subsection{peak separation error}
\begin{figure*}
\begin{center}
\includegraphics[width=1.3\columnwidth,clip]{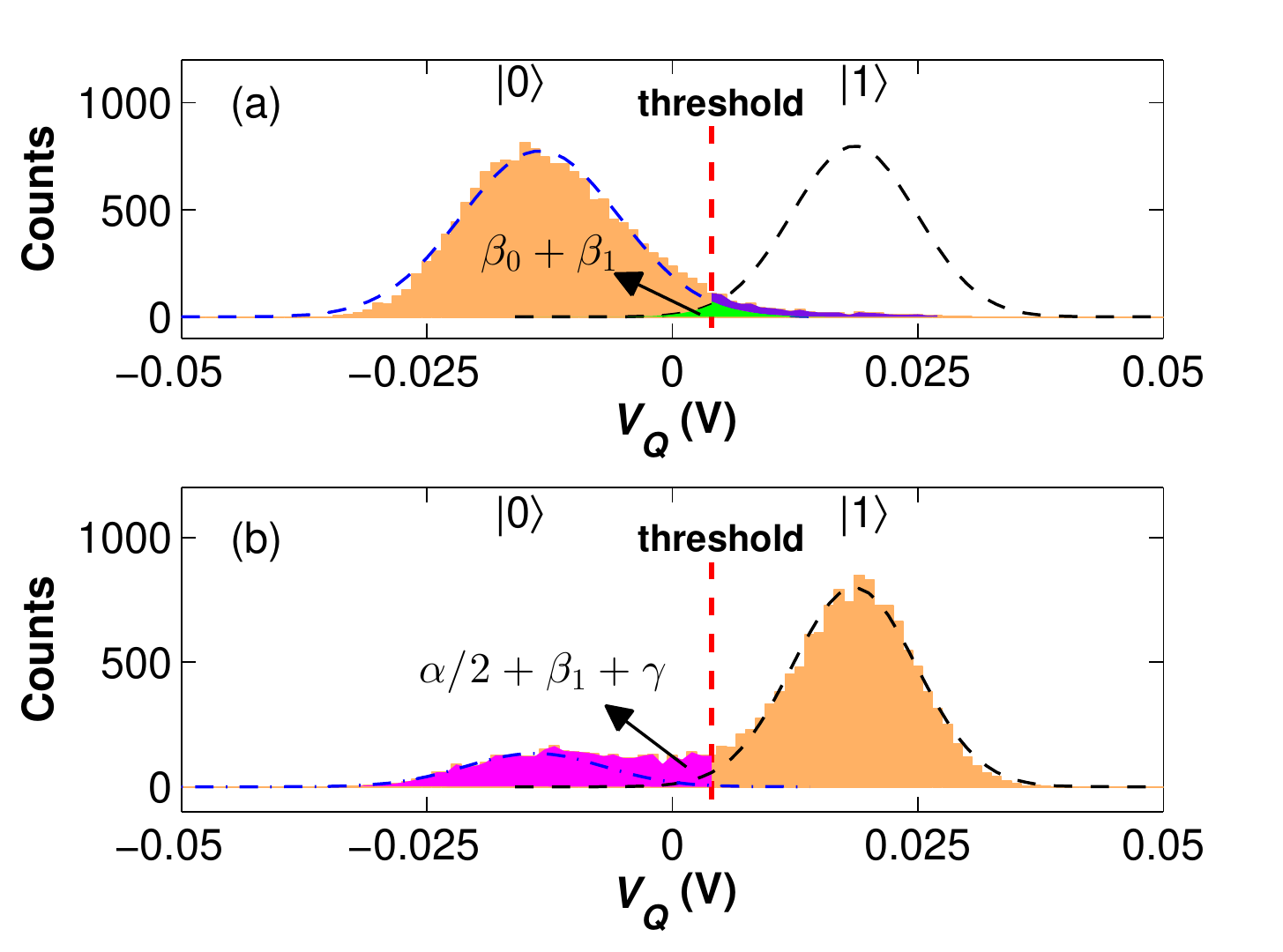}
\end{center}
\caption{Histograms obtained when the $\pi$ pulse is (a) turned off and (b) turned on. 
The condition of the measurement is the same as that used for Fig.~3(e) in the main article. 
Namely, the JPA is operated at a gain of 25.6~dB and a bandwidth of 4.8~MHz, 
and the delay time $t_{\rm d}$ is 100~ns. 
The red dashed line is the threshold used to discriminate the qubit states.
The blue dashed curve in (a) is a Gaussian fit to the data below threshold, 
while the black dashed curve in (b) is a Gaussian fit to the data above the threshold. 
The black dashed curve in (a) is the same as the one in (b).
The blue dash-dot curve in (b) is a gaussian fit to the data for $V_Q \le V_Q^0$, 
where $V_Q^0$ is the center of the fitting curve for $|0\rangle$-state peak in (a). 
}
\label{figS2}
\end{figure*}

Figures~\ref{figS2}(a) and (b) show the histograms 
when the $\pi$ pulse is turned off and turned on, respectively. 
The condition of the measurement is the same as that used for Fig.~3(e) in the main article. 
Namely, the JPA is operated at a gain of 25.6~dB and a bandwidth of 4.8~MHz, 
and the delay time $t_{\rm d}$ is 100~ns. 
In Fig.~\ref{figS2}(a), we observe one main peak corresponding
to the qubit state $|0\rangle$.
In Fig.~\ref{figS2}(b), on the other hand, 
in addition to a main peak corresponding to the qubit state $|1\rangle$, 
we observe counts below the threshold which are mainly due to qubit energy relaxation 
as discussed below. 

We fit the main peaks by a Gaussian function using the data below (above) the 
threshold in Fig.~\ref{figS2}(a) [Fig.~\ref{figS2}(b)]. 
The error due to insufficient separation of the peaks in the histogram
is estimated by calculating the overlap between the two Gaussian curves 
(blue and black dashed curves).
The overlap is shown by the green area in Fig.~\ref{figS2}(a).
The area below the threshold gives the discrimination error $\beta_1 =0.9\%$
for state $|1\rangle$, while 
the area above the threshold gives the discrimination error of $\beta_0 =1.5\%$
for state $|0\rangle$.
In total, we estimate the peak separation error $\beta$ to be 2.4\%. 
Note that the $|0\rangle$-sate peak in Fig.~\ref{figS2}(a) is 
quite asymmetric and different from the $|1\rangle$-state peak in Fig.~\ref{figS2}(b). 
We will discuss this point later. 

\subsection{energy relaxation of the qubit}
In Fig.~\ref{figS2}(b), we observe counts below the threshold which are mainly due to
qubit energy relaxation. 
Since our readout acquisition time is the same as the delay time $t_{\rm d}$, 
we cannot neglect qubit relaxation during the acquisition. 
Roughly speaking, qubit relaxation during the delay time is manifest 
as a peak with the same mean value and width as 
those of the state $|0\rangle$ peak in Fig.~\ref{figS2}(a), 
while the relaxation during the acquisition time is manifest 
as counts between the main peaks corresponding to the states $|0\rangle$ and $|1\rangle$. 
The blue dash-dot curve is a Gaussian fit to the data for $V_Q \le V_Q^0$, 
where $V_Q^0$ is the center of the fitting curve 
for state $|0\rangle$ in Fig.~\ref{figS2}(a).
The area of the blue dash-dot curve equals 14.8\% of the total counts $N_{\rm tot}$, 
which is close to what is expected from the qubit energy relaxation during $t_{\rm d}$, 
namely, $1-\exp(-t_{\rm d}/T_1)=15.1\%$. 
Here, $T_1 = 610$~ns is the mean value 
for $\langle n\rangle \le 1.7$ shown in Fig.~2(a) in the main article. 
The deviation of the data from the fitting curve around $V_Q=0$ is 
attributable to the qubit energy relaxation during the acquisition. 
The magenta area represents the total error when a $\pi$-pulse is turned on. 
It includes $\alpha/2$, $\beta_1$, 
and the error due to energy relaxation of the qubit $\gamma$. 
Based on this, we estimate $\gamma$ to be 19.8\%. 

\subsection{unaccounted error}
\begin{figure*}
\begin{center}
\includegraphics[width=1.3\columnwidth,clip]{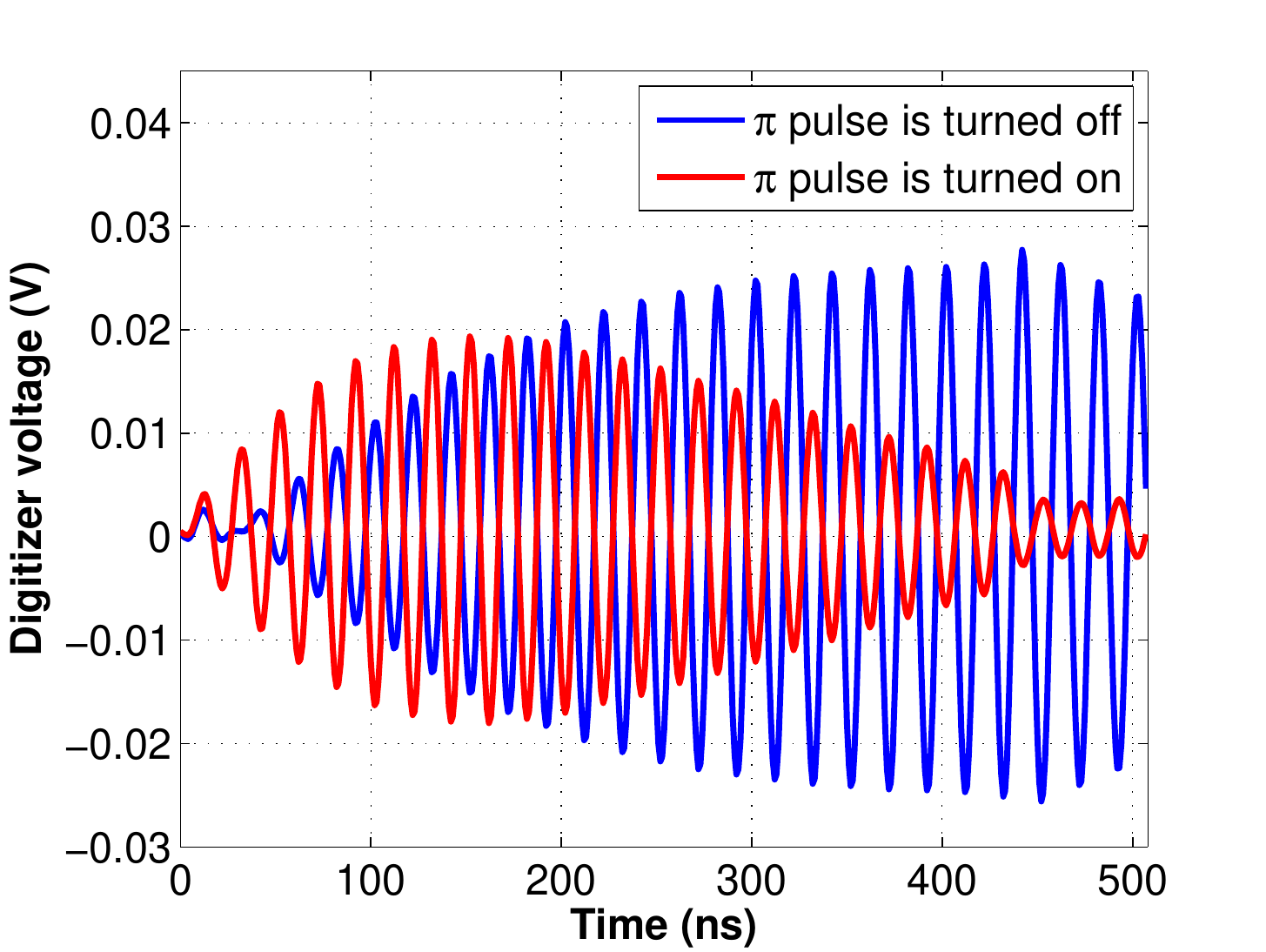}
\end{center}
\caption{
Timetraces of the heterodyne signal (IF frequency 50~MHz) 
when the $\pi$ pulse is turned off (blue) and turned on (red). 
The data is averaged $6.5\times 10^4$ times for each measurement point. 
The JPA is operated with a gain of 29~dB and a bandwidth of 2.8~MHz.
}
\label{figS3}
\end{figure*}

The three errors described above sum to 25.0\%, 
and we still have an unaccounted error of 1.5\%. 
The histograms in Fig.~\ref{figS2} exhibit an asymmetry. 
In particular, the state $|0\rangle$ distribution in Fig.~\ref{figS2}(a) 
has a tail toward higher $V_Q$. 
The total counts in the purple area in Fig.~\ref{figS2}(a) 
exceed $N_{\rm tot}\gamma/2$, 
and this accounts for the remaining error. 
One reason for the tail is the transient response of JPA. 
Namely, we start the acquisition of the readout voltage pulse 
before its amplitude fully saturates. 
Actually, at larger $t_{\rm d}$ or smaller $P_{\rm p}$ (larger bandwidth), 
this asymmetry becomes less conspicuous. 
Note that the $|1\rangle$-state peak in Fig.~\ref{figS2}(b) 
is less asymmetric and the peak width is smaller. 
One reason for this is the difference in the response of the readout pulse for the two qubit states. 
As exemplified in Fig.~\ref{figS3}, when the qubit is in state $|0\rangle$, 
there is $\sim40$~ns delay before the voltage starts to rise up, 
compared to the case when the qubit is in $|1\rangle$ state. 
Using a time constant $\tau$ of 66~ns for 4.8~MHz bandwidth of the JPA 
and $t_{\rm d}$ of 100~ns, 
$1-\exp(-t_{\rm d}/\tau)$ is 0.78, while $1-\exp[-(t_{\rm d}-40~\rm ns)/\tau)]$ is 0.60. 
This qualitatively explains the smaller effect of the transient response of the JPA 
for the case when the $\pi$ pulse is turned on. 

\section{Noise temperature of the JPA}
\begin{figure*}
\begin{center}
\includegraphics[width=1.3\columnwidth,clip]{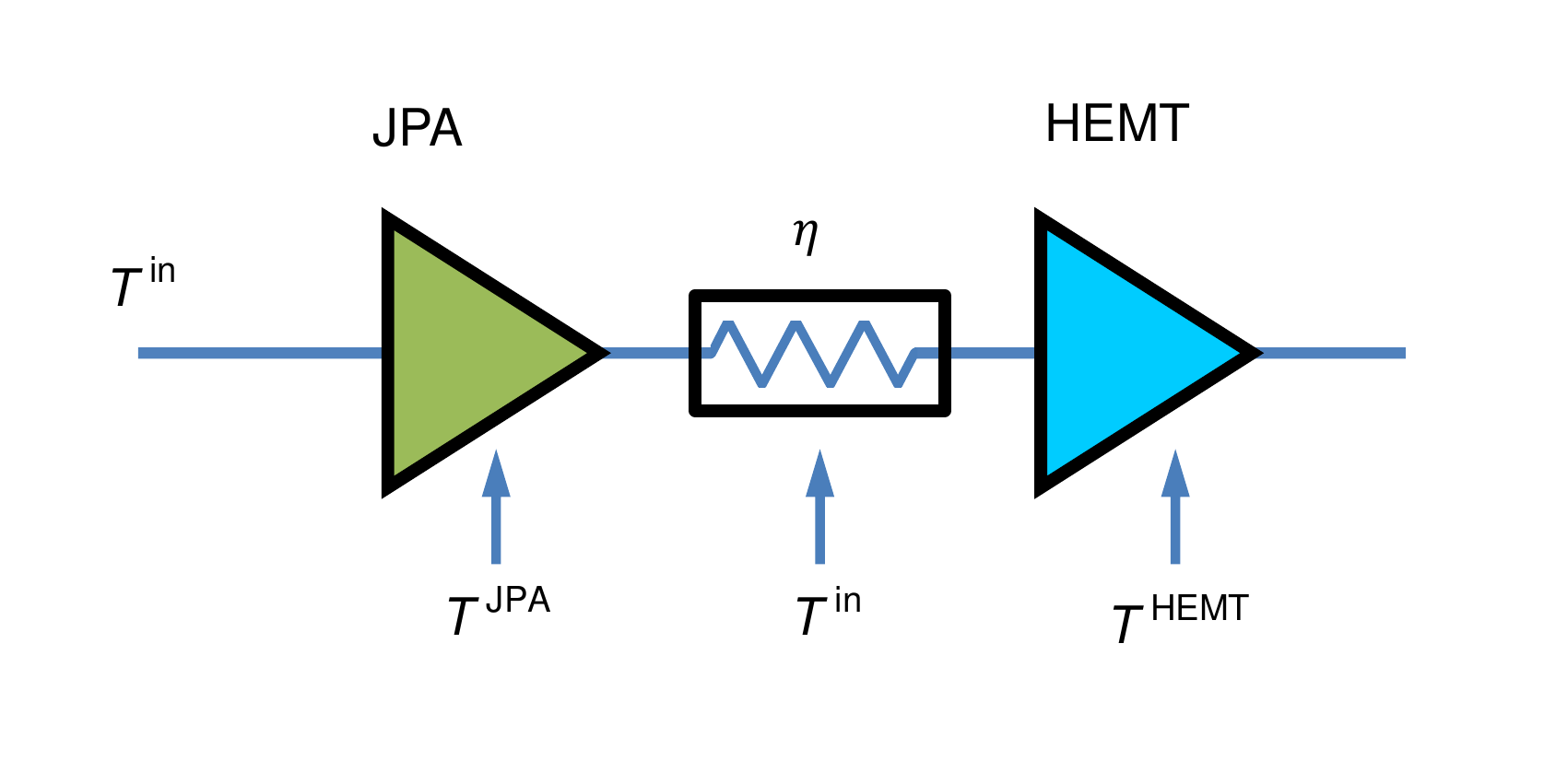}
\end{center}
\caption{Simplified diagram for estimating the noise temperature of the JPA.
}
\label{figS4}
\end{figure*}

In the main article, we showed that the upper limit of the
noise temperature of the JPA is 380 mK. Here, we explain this. 
We consider the chain of amplifiers as shown in Fig.~\ref{figS4}, in which we have JPA, HEMT and the
loss between them.
The noise power referred to the input of the HEMT is given by
\begin{equation}
P_{\rm noise} = [(1- \eta) k_{\rm B}(T^{\rm in}+T^{\rm JPA})G_{\rm eff}
+ \eta k_{\rm B}T^{\rm in} + k_{\rm B}T^{\rm HEMT}]\Delta f, \label{NT}
\end{equation}
where $k_{\rm B}$ is the Boltzmann constant,
$T^{\rm JPA}$ and $T^{\rm HEMT}$ are the noise temperature of the JPA and the HEMT, respectively,
$T^{\rm in}$ represents the input noise from the 50 Ohm resistor, 
$\Delta f$ is the measurement bandwidth,
$\eta$ is the loss between the JPA and the HEMT,
and $G_{\rm eff}$ is the effective JPA gain for white uncorrelated noise.  

We have not measured $T^{\rm HEMT}$ precisely,
and assume it to be $6$ K based on the catalog specification measured at $15$ K. 
$T^{\rm in}$ is assumed to be the vacuum noise, namely, $k_{\rm B}T^{\rm in} = h f_0/2$,
where $h$ is the Planck constant and $f_0$ is the signal frequency.
The loss represented by $\eta$ is due to that in the connectors, circulators, filters, and isolators
(which are in the base temperature stage)
between JPA and HEMT, and $\eta$ is estimated to be $< 2$ dB
from independent measurements. $\Delta f$ is $10$ MHz which corresponds to the acquisition time $\tau$ of 100~ns.
$G_{\rm eff}$ is given by $\frac{1}{\Delta f}\int_{-\infty}^{\infty}[G_{\rm s}(f)+G_{\rm i}(f)]{\rm sinc}^2(2\pi f\tau/2) df$,
where $G_{\rm s}(f)$ and $G_{\rm i}(f)$ represent signal and idler gains as functions of the frequency, respectively. 
Their Lorentzian linewidths are measured to be $4.4$ MHz and the maximum gains, $G_{\rm s}(f_0)$ and $G_{\rm i}(f_0)$ are 20.3 dB. 

From the histogram for state-$|1 \rangle$ peak at $t_{\rm d} = 200$ ns, we obtain the signal to 
noise ratio
$ {\rm SNR} = \mu_1 /\sigma _1 \approx $ 2.9,
where $\mu_1$ and $\sigma_1$ are the  mean value and the  standard deviation for the Gaussian fit to state-$|1 \rangle$ peak.
By using Eq. (\ref{NT}) and  $ {\rm SNR} = \sqrt{P_{\rm s} / P_{\rm noise}}$,
where $P_{\rm s}$ is the signal power input to the HEMT and estimated to be $-105 \pm 2$ dBm, 
we get $T^{\rm JPA} < 380 $ mK.  
Because $T^{\rm HEMT}/[G_{\rm eff}(1-\eta)]$ is $< 100$ mK,
the input vacuum noise and possibily the noise temperature of the JPA are mainly limiting the SNR.


\begin{thebibliography}{27}
\bibitem[{com()}]{comment1}
\bibinfo{note}{There is an apparent avoided-level crossing around
  $\Phi/\Phi_0=0.29$. This is an artifact because of the 2$\pi$ span of the
  color chart, while there is 4$\pi$ rotation in total, namely $2\pi$ from each
  resonator.}
\end{thebibliography}

\end{document}